\documentclass[twocolumn,amsmath,amssymb]{revtex4}
\usepackage{amsfonts}
\usepackage[usenames]{color}

\newcommand{\be}{\begin{equation}}
\newcommand{\ee}{\end{equation}}
\newcommand{\bea}{\begin{eqnarray}}
\newcommand{\eea}{\end{eqnarray}}

\begin{document}
\title{Definition of electric and magnetic fields in curved spacetime}
\author{Jai-chan Hwang${}^{1}$, Hyerim Noh${}^{2}$}
\address{${}^{1}$Particle Theory  and Cosmology Group,
         Center for Theoretical Physics of the Universe,
         Institute for Basic Science (IBS), Daejeon, 34126, Republic of Korea
         \\
         ${}^{2}$Theoretical Astrophysics Group, Korea Astronomy and Space Science Institute, Daejeon, Republic of Korea
         }


\begin{abstract}

Defining the electric and magnetic field vectors in curved spacetime requires a proper choice of the observer's frame four-vector. Related literature shows that this fundamental issue in physics still needs to be properly resolved. In recent literature on using electromagnetic means to detect gravitational waves, an {\it ad hoc} definition based on regarding $F_{ab}$ with two covariant indices as the special relativistic one is popular. We show that by assigning physical fields to tensor components in that way, we cannot identify the frame four-vector allowing such a choice, thus failing to properly define the external charge and current densities in that frame. We propose the normal frame as the proper one. In this frame, the weak gravity corrections appear as the effective polarizations and magnetizations in both the homogeneous and inhomogeneous parts of Maxwell's equations.

\end{abstract}

\maketitle

%
%
%
\section{Introduction}

The definition of the electric and magnetic fields, {\bf E} and {\bf B}, depends on the observer's motion. The advent of special relativity by Einstein made this point clear \cite{Einstein-1905}. The four-dimensional extension of Maxwell's equations was made by Minkowski \cite{Minkowski-1910} by introducing electromagnetic (EM) field strength tensor $F_{ab}$; for a historical note on the development, see Pauli in Sec.\ 28 of \cite{Pauli-1958}. In special relativity, the relation between the EM fields and the field strength tensor is trivial: one can simply identify components of $F_{ab}$ to corresponding {\bf E} and {\bf B}. The case is not that simple in curved spacetime, and the correspondence depends on gravity as well.

After introducing general relativity, Einstein suggested the four-dimensional Maxwell's equations of Minkowski, now in curved spacetime, by using interesting coincidences he found in a couple of similar identifications directly made in $F_{ab}$ and its duel tensor \cite{Einstein-1916}; the result is the four-dimensional Maxwell's equations we know currently, Eq.\ (\ref{Maxwell-tensor-eqs}). We wish to stress that Einstein used such identifications only to reach the correct form of four-dimensional Maxwell's equations in curved spacetime, and he has {\it not} defined the EM fields in that way. In a curved spacetime, not only are the indices of $F_{ab}$ raised and lowered by the spacetime metric but $F_{ab}$ with two covariant indices can also depend on the metric tensor. We will come up with further serious trouble of directly matching the fields to components of a tensor $F_{ab}$, see also \cite{Crater-1994}. Then, how do we define the EM fields in curved spacetime?

The interaction of spacetime metric with electromagnetism introduces novel ways of detecting gravitational waves in the laboratory. For this purpose, a proper definition of EM fields is a prerequisite. In the related literature, however, the electric and magnetic fields in curved spacetime metric are often defined in an {\it ad hoc} way assuming $F_{ab}$ (with two covariant indices) keeps special relativistic form \cite{Cooperstock-1968, Baroni-1985, Berlin-2022, Domcke-2022}. In this way, the weak gravity corrections in Maxwell's equations appear as the effective polarization and magnetization vectors, ${\bf P}$ and ${\bf M}$, or effective charge and current densities, only in the inhomogeneous part. This has a practical advantage and the situation is analogous to the helically coupled electromagnetism due to axion \cite{Sikivie-1983, Wilczek-1987}, thus with similar detection mechanism proposed in \cite{Berlin-2022, Domcke-2022}.

However, the arbitrary nature of this choice can be shown as we can similarly choose its duel $F^*_{ab}$ in a special relativistic form, as Einstein did, with consequent effective ${\bf P}$ and ${\bf M}$ now appearing only in the homogeneous part. We can similarly make many different definitions with less striking consequences. As mentioned, these two choices were considered by Einstein in his way of guessing the proper tensorial form of Maxwell's equations in general relativity \cite{Einstein-1916}, but he has {\it not} used those to define the fields.

A covariant way of decomposing $F_{ab}$ into the EM fields is known \cite{Moller-1952, Lichnerowicz-1967, Ellis-1973}. In the covariant decomposition, choosing an observer's frame four-vector corresponds to defining the EM field vectors covariantly. In the literature, the comoving and the normal frames are often used \cite{Hwang-Noh-2022-Axion-EM},
and the normal frame is exclusively used in the numerical relativity \cite{Baumgarte-Shapiro-2010, Gourgoulhon-2012, Shibata-2015, Baumgarte-Shapiro-2021}. In the normal frame, we will show that the weak gravity causes effective ${\bf P}$s and ${\bf M}$s appearing in {\it both} the homogeneous and inhomogeneous parts of Maxwell's equations.

If we set aside the observational identification of proper fields out of these three or many other choices, any choice is fine, as the fields in different frames are related to each other algebraically. With the former two non-covariant choices, however, it happens that we {\it cannot} identify the frame four-vectors even in the perturbation analyses. Consequently, the external charge and current densities, which are similarly frame dependent, cannot be identified/defined in these two frames. This is troublesome both mathematically and physically.

Here, in the case of weak gravity, we derive Maxwell's equations in the above three different definitions and express the gravity-caused corrections as effective ${\bf P}$s and ${\bf M}$s. We propose the normal frame as the proper choice based on the nature of the normal frame. In Appendix \ref{sec:coordinate} we present coordinate frame where the observer is attached to the spatial coordinate. The Appendix \ref{sec:GT} presents the gauge-invariance of the four sets of Maxwell's equations expanded to the second-order perturbation. 

%
%
%
\section{Frame dependence}

In terms of the EM field strength tensor $F_{ab}$, Maxwell's equations, in Heaviside unit \cite{Jackson-1975}, are
\bea
   F^{ab}_{\;\;\;\;;b}
       = {1 \over c} J^a, \quad
       \eta^{abcd} F_{bc,d} = 0.
   \label{Maxwell-tensor-eqs}
\eea
The first equation gives the inhomogeneous part of Maxwell's equations (Gauss' and Amp$\grave{\rm e}$re's laws), and the second one gives homogeneous part (no-monopole constraint and Faraday's law). Using its duel, $F^*_{ab} \equiv {1 \over 2} \eta_{abcd} F^{cd}$, Eq.\ (\ref{Maxwell-tensor-eqs}) becomes
\bea
   - {1 \over 2} \eta^{abcd} F^*_{bc,d}
       = {1 \over c} J^a,
       \quad
       F^{*ab}_{\;\;\;\;\;\;;b} = 0.
   \label{Maxwell-tensor*-eqs}
\eea
Maxwell's equations in four-dimensional form were first introduced by Minkowski in special relativity \cite{Minkowski-1910} and by Einstein in general relativity \cite{Einstein-1916}. These equations are covariant and valid independently of the frame and gauge choice (in both gravity and EM sectors).

To introduce the EM fields, we need to introduce the frame four-vector related to the observer. In terms of a generic time-like four-vector $U_a$ with $U_a U^a \equiv -1$, we can introduce these fields in a covariant way as \cite{Moller-1952, Lichnerowicz-1967, Ellis-1973}
\bea
   & & F_{ab} \equiv U_a E_b - U_b E_a
       - \eta_{abcd} U^c B^d,
   \nonumber \\
   & & F^*_{ab} = U_a B_b - U_b B_a
       + \eta_{abcd} U^c E^d.
   \label{Fab-cov}
\eea
For an observer moving with the four-velocity $U_a$, the electric and magnetic fields are
\bea
   E_a = F_{ab} U^b, \quad
       B_a = F_{ab}^* U^b,
\eea
with $E_a U^a \equiv 0 \equiv B_a U^a$. The current four-vector is decomposed using the {\it same} four-vector as
\bea
   J^a \equiv \varrho c U^a
       + j^a, \quad
       j_a U^a \equiv 0,
   \label{J^a}
\eea
where $\varrho$ and $j_a$ are charge and current densities, respectively. Choosing the frame, corresponding to choosing the observer's four-velocity, has nothing to do with the gauge condition, and all our arguments are valid without imposing the gauge (coordinate) conditions. The covariant form of Maxwell's equations associated with the generic frame four-vector are derived in \cite{Ellis-1973, Hwang-Noh-2022-Axion-EM}.

Popular frame vectors used in the literature are the comoving frame with $u_a$ the accompanying fluid (or field) four-vector, and the normal frame $n_a$ with $n_i \equiv 0$; in the presence of multiple fluids and fields, we have many different choices for the comoving frame, and in the normal frame case the fluid (or field) velocity component appears as the flux term. In the following, we reserve $E_a$, $B_a$, $\varrho$ and $j_a$ for EM fields and charge and current densities observed by the normal observer with $n_a$ as the observer's four-velocity.

We will derive Maxwell's equations to the second-order perturbations in Minkowski background. For consistency, we regard the presence of EM fields and four-current as perturbations. The additional presence of metric perturbation may cause changes in the EM field which can be regarded as second-order perturbations. As Maxwell's equations are linear in the EM fields (charge and current densities as well), in order to derive the metric-caused changes in the EM field, we need metric perturbation only to the linear order.

We consider the most general linear perturbations in Minkowski background
\bea
   & & g_{ab} \equiv \eta_{ab} + h_{ab}, \quad
       g^{ab} = \eta^{ab} - h^{ab},
   \nonumber \\
   & & \Gamma^a_{bc} = {1 \over 2} ( h^a_{b,c}
       + h^a_{c,b} - h_{bc}^{\;\;\;,a} ),
   \nonumber \\
   & & R^a_{\;\;bcd} = {1 \over 2} ( h^a_{d,bc}
       + h_{bc\;\;\;d}^{\;\;\;\,,a} - h^a_{c,bd}
       - h_{bd\;\;\;c}^{\;\;\;\;,a} ).
   \label{metric-pert}
\eea
As we consider linear perturbation in the metric, indices of $h_{ab}$ can be raised and lowered using $\eta_{ab}$ and its inverse. The gauge conditions are not imposed, and we can show that all the Maxwell's equations in this work are gauge-invariant to the second-order perturbation, see the Appendix \ref{sec:GT}. We have
\bea
   & & \eta_{0ijk} = - \sqrt{-g} \eta_{ijk}, \quad
       \eta^{0ijk} = {1 \over \sqrt{-g}} \eta^{ijk},
   \nonumber \\
   & &
       g \equiv {\rm det}(g_{ab}) = - ( 1 + h^0_0 + h^i_i ),
\eea
where indices of $\eta_{ijk}$ are raised and lowered using $\delta_{ij}$ and its inverse; $i, j \dots$ are spatial indices.

In the following we consider three different ways of defining the EM fields. The coordinate frame is presented in Appendix \ref{sec:coordinate}.

\subsection{Normal frame}
                                  \label{normal-frame}

For the EM fields and current density in the normal frame, we set
\bea
   \widetilde B_i \equiv B_i, \quad
       \widetilde B_0 = h^i_0 B_i, \quad
       \widetilde B^i = B^i - h^{ij} B_j, \quad
       \widetilde B^0 = 0,
\eea
etc.; indices of $B_i$, $E_i$ and $j_i$ are raised and lowered using $\delta_{ij}$ and its inverse; we often use an overtilde like $\widetilde B_a$ to indicate a covariant quantity. In the normal frame with
\bea
   \widetilde n_i \equiv 0, \quad
       \widetilde n_0 = - 1 - {1 \over 2} h^0_0, \quad
       \widetilde n^i = - h^i_0, \quad
       \widetilde n^0 = 1 - {1 \over 2} h^0_0,
   \label{n_a-ADM}
\eea
Eqs.\ (\ref{Fab-cov}) and (\ref{J^a}) give
\bea
   & & \hskip -.9cm
       \widetilde F_{0i} = - \left( 1 + {1 \over 2} h^0_0 \right) E_i
       - \eta_{ijk} h^j_0 B^k,
   \nonumber \\
   & & \hskip -.9cm
       \widetilde F_{ij} = \eta_{ijk} \left[ \left( 1
       + {1 \over 2} h^\ell_\ell \right) B^k
       - h^{k\ell} B_\ell \right],
   \nonumber \\
   & & \hskip -.9cm
       \widetilde J^0
       = \varrho c
       \left( 1 - {1 \over 2} h^0_0 \right), \quad
       \widetilde J^i = j^i - h^{ij} j_j
       - \varrho c h^i_0.
   \label{Fab-normal}
\eea
Thus, in the normal frame $F_{ab}$ is dependent on the metric perturbations.

Equation (\ref{Maxwell-tensor-eqs}), with $x^0 = ct$, gives
\bea
   & & ( E^i + P_{\rm E}^i )_{,i}
       = \varrho \left( 1 + {1 \over 2} h^i_i \right)
       \equiv \overline \varrho,
   \label{Maxwell-normal-1} \\
   & &
       ( E^i + P_{\rm E}^i )_{,0}
       - \eta^{ijk} \nabla_j
       ( B_k- M^{\rm E}_k )
   \nonumber \\
   & & \qquad
       =
       - {1 \over c} \left( 1 + {1 \over 2} h^j_j
       + {1 \over 2} h^0_0 \right) j^i
       + {1 \over c} h^{ij} j_j
       + \varrho h^i_0
   \nonumber \\
   & & \qquad
       \equiv - {1 \over c} \overline j^i,
   \label{Maxwell-normal-2} \\
   & & ( B^i + P_{\rm B}^i )_{,i} = 0,
   \label{Maxwell-normal-3} \\
   & & ( B^i + P_{\rm B}^i )_{,0}
       + \eta^{ijk} \nabla_j ( E_k - M^{\rm B}_k )
       = 0,
   \label{Maxwell-normal-4}
\eea
where the effective ${\bf P}$s and ${\bf M}$s caused by the metric are
\bea
   & & \hskip -1cm P_{\rm E}^i
       \equiv {1 \over 2} h^j_j E^i - h^{ij} E_j, \quad
       M_{\rm E}^i
       \equiv {1 \over 2} h_{00} B^i + \eta^{ijk} h_{0j} E_k,
   \nonumber \\
   & & \hskip -1cm P_{\rm B}^i
       \equiv {1 \over 2} h^j_j B^i - h^{ij} B_j, \quad
       M_{\rm B}^i
       \equiv {1 \over 2} h_{00} E^i - \eta^{ijk} h_{0j} B_k.
   \label{PM-normal}
\eea
Therefore, in the normal frame, the weak gravity causes the effective ${\bf P}$s and ${\bf M}$s in both the homogeneous and inhomogeneous parts of Maxwell's equations, or the effective charge and current densities in both parts as $\varrho_{\rm E} \equiv - P_{{\rm E},i}^i$ and $j_{\rm E}^i \equiv \dot P_{\rm E}^i + c \eta^{ijk} \nabla_j M^{\rm E}_k$, and similarly for $\varrho_{\rm B}$ and $j_{\rm B}^i$.


\subsection{Special relativistic $F_{ab}$}

Without resorting to the frame four-vector, the EM fields are often defined directly from components of the EM field strength tensor. This is casual in special relativity but troublesome in curved spacetime. Still, a popular one in the literature related to gravitational wave detection using EM means is setting $F_{ab}$ with two covariant indices in the special relativistic form \cite{Cooperstock-1968, Baroni-1985, Berlin-2022, Domcke-2022}; the same definition is also suggested in Sec.\ 90 of \cite{Landau-Lifshitz-1975} which is the only textbook attempting the definition in curved spacetime; referring to Eq.\ (\ref{Maxwell-electric-3-4}), the authors add ``the analogy (purely formal, of course)" though. Thus, the EM fields are defined as
\bea
   \widetilde F_{ij}
       \equiv \eta_{ijk} {\hat B}^k, \quad
       \widetilde F_{0i}
       \equiv - {\hat E}_i,
   \label{Fab-electric}
\eea
where indices of ${\hat E}_i$ and ${\hat B}_i$ are raised and lowered using $\delta_{ij}$ and its inverse; for clarity we use different notations for the EM fields. In this way, the second part of Eq.\ (\ref{Maxwell-tensor-eqs}) naturally gives the homogeneous part of Maxwell's equations in exactly special relativistic form
\bea
   {\hat B}^i_{\;\;,i} = 0, \quad
       {\hat B}^i_{\;\;,0}
       + \eta^{ijk} \nabla_j {\hat E}_k = 0.
   \label{Maxwell-electric-3-4}
\eea
Einstein derived these equations for some other purpose \cite{Einstein-1916}. In terms of ${\hat E}_i$ and ${\hat B}_i$, the other Maxwell's equations in exact forms become highly complicated. The trouble with this non-covariant way of assigning physical vectors to the component of the field strength tensor is that, let alone its ambiguous physical meaning \cite{Crater-1994}, we {\it cannot} identify the (observer's) frame four-vector $\hat U_a$ allowing such a choice.

To the second order, from Eqs.\ (\ref{Fab-normal}) and (\ref{Fab-electric}), we have
\bea
   & & \hat E_i = \left( 1 + {1 \over 2} h^0_0 \right) E_i
       + \eta_{ijk} h_0^j B^k
       = E_i - M_i^{\rm B},
   \nonumber \\
   & & \hat B_i = \left( 1 + {1 \over 2} h^j_j \right) B_i
       - h_{ij} B^j
       = B_i + P_i^{\rm B}.
   \label{EB-electric}
\eea
Rather unexpectedly, even to the second order perturbation we cannot identify a frame four-vector $\hat U_a$ accompanied by the choice in Eq.\ (\ref{Fab-electric}). With hindsight, though, the result is understandable because we only have three degrees of freedom for the normalized four-vector whereas the relation between $F_{ab}$ and the two EM fields has six independent degrees of freedom; in general, one cannot satisfy six relations only with three conditions. Using Eq.\ (\ref{EB-electric}), Eqs.\ (\ref{Maxwell-normal-1})-(\ref{PM-normal}) give
\bea
   & & ( {\hat E}^i + {\hat P}_{\rm E}^i )_{,i}
       = \overline \varrho,
   \label{Maxwell-electric-1} \\
   & &
       ( {\hat E}^i + {\hat P}_{\rm E}^i )_{,0}
       - \eta^{ijk} \nabla_j ( {\hat B}_k
       - {\hat M}^{\rm E}_k )
       = - {1 \over c} \overline j^i,
   \label{Maxwell-electric-2} \\
   & & {\hat B}^i_{\;\;,i} = 0,
   \label{Maxwell-electric-3} \\
   & & {\hat B}^i_{\;\;,0}
       + \eta^{ijk} \nabla_j {\hat E}_k = 0,
   \label{Maxwell-electric-4}
\eea
where
\bea
   & & {\hat P}_{\rm E}^i
       \equiv {1 \over 2} \left( h^j_j - h^0_0 \right) {\hat E}^i
       - h^{ij} {\hat E}_j
       - \eta^{ijk} h_{0j} {\hat B}_k,
   \nonumber \\
   & & {\hat M}_{\rm E}^i
       \equiv {1 \over 2} \left( h^j_j - h^0_0 \right) {\hat B}^i
       - h^{ij} {\hat B}_j
       + \eta^{ijk} h_{0j} {\hat E}_k.
   \label{PM-electric}
\eea
These coincide with results in \cite{Cooperstock-1968, Baroni-1985, Berlin-2022, Domcke-2022}.

The presence of effective ${\bf P}$ and ${\bf M}$ only in the inhomogeneous parts of Maxwell's equations is an important advantage. However, notice that for the charge and current densities, we kept the ones in the normal frame; without identifying the frame four-vector for the choice in Eq.\ (\ref{Fab-electric}), we {\it cannot} define the charge and current densities in that frame. The lack of associated charge and current densities is a serious problem.

Next, we point out the {\it ad hoc} nature of this definition by showing a similar choice with an opposite result.

\subsection{Special relativistic $F^*_{ab}$}

Now, we take $F^*_{ab}$ the special relativistic form, i.e., independent of metric perturbations. Thus, the EM fields, now using another notation ${\breve E}_i$ and ${\breve B}_i$, are defined as
\bea
   \widetilde F^*_{ij}
       \equiv - \eta_{ijk} {\breve E}^k, \quad
       \widetilde F^*_{0i}
       \equiv - {\breve B}_i,
   \label{Fab-magnetic}
\eea
where indices of ${\breve E}_i$ and ${\breve B}_i$ are raised and lowered using $\delta_{ij}$ and its inverse. In this way, in the absence of the external sources, the first part of Eq.\ (\ref{Maxwell-tensor*-eqs}) naturally gives the inhomogeneous part of Maxwell's equations in the exactly special relativistic form
\bea
   {\breve E}^i_{\;\;,i} = 0, \quad
       {\breve E}^i_{\;\;,0}
       - \eta^{ijk} \nabla_j {\breve B}_k = 0.
   \label{Maxwell-magnetic-3-4}
\eea
In the presence of the sources, the absence of the frame four-vector makes the external source difficult to define. Einstein derived these equations with external sources, but by simply assuming such sources exist formally \cite{Einstein-1916}; but his purpose was not for suggesting this or the previous definitions in Eq.\ (\ref{Fab-electric}) for the EM fields. Using ${\breve E}_i$ and ${\breve B}_i$, the other Maxwell's equations become highly complicated. According to Einstein, two vectors ${\hat E}_i$ and ${\hat B}_i$ are in pretty complex relationship with ${\breve E}_i$ and ${\breve B}_i$ which is determined by $F^*_{ab} = {1 \over 2} \eta_{abcd} g^{ce} g^{df} F_{ef}$ \cite{Einstein-1916}.

To the second order, from Eqs.\ (\ref{Fab-normal}) and (\ref{Fab-magnetic}), we have
\bea
   & & {\breve E}_i = \left( 1 + {1 \over 2} h^j_j \right) E_i
       - h_i^j E_j
       = E_i + P_i^{\rm E},
   \nonumber \\
   & & {\breve B}_i = \left( 1 + {1 \over 2} h^0_0 \right) B_i
       - \eta_{ijk} h_0^j E^k
       = B_i - M_i^{\rm E}.
   \label{EB-magnetic}
\eea
Here again, we cannot identify a frame four-vector $\breve U_a$ accompanied by the choice in Eq.\ (\ref{Fab-magnetic}). In terms of ${\breve E}_i$ and ${\breve B}_i$, but keeping the charge and current densities in the normal frame ones, Eqs.\ (\ref{Maxwell-normal-1})-(\ref{PM-normal}) lead to
\bea
   & & {\breve E}^i_{\;\;,i}
       = \overline \varrho,
   \label{Maxwell-magnetic-1} \\
   & &
       {\breve E}^i_{\;\;,0}
       - \eta^{ijk} \nabla_j {\breve B}_k
       = - {1 \over c} \overline j^i,
   \label{Maxwell-magnetic-2} \\
   & & ( {\breve B}^i + {\breve P}_{\rm B}^i )_{,i} = 0,
   \label{Maxwell-magnetic-3} \\
   & & ( {\breve B}^i + {\breve P}_{\rm B}^i )_{,0}
       + \eta^{ijk} \nabla_j ( {\breve E}_k
       - {\breve M}^{\rm B}_k ) = 0,
   \label{Maxwell-magnetic-4}
\eea
where $\breve P_{\rm B}^i = \hat M_{\rm E}^i$ and $\breve M_{\rm B}^i = \hat P_{\rm E}^i$. The effective ${\bf P}$ and ${\bf M}$ now appear in the homogeneous part of Maxwell's equations which is opposite to the previous choice.

\section{Discussion}
                                   \label{sec:Discussion}

We derived Maxwell's equations in curved spacetime with the most general linear metric perturbations in the Minkowski background. The EM fields depend on the choice of the observer (the frame four-vector); in the covariant expression $F_{ab}$ and $J_a$ are independent of the frame (observer), but as their decomposition, $E_a$ and $B_a$ as well as $\varrho$ and $j_a$, are frame dependent. Thus, a proper choice of the frame is essential for handling Maxwell's equations in curved spacetime.

In the literature concerned with detecting gravitational waves by EM means, a special definition without resorting to the frame four-vector is popular where the EM fields are introduced assuming $F_{ab}$ is the one in Minkowski space \cite{Cooperstock-1968, Baroni-1985, Berlin-2022, Domcke-2022}. Although it is an {\it ad hoc} choice, by itself it is not a problem as we can simply translate the fields from one frame to another.

Here, however, we point out the arbitrary nature of the choice and also the trouble of this non-covariant definition. The former point is indicated by comparing with the opposite result coming from similarly introduced fields based on assuming $F^*_{ab}$ as the special relativistic one. The latter point is related to the absence of the frame four-vector $U_a$ enabling these two definitions. The absence of the frame four-vector is a serious drawback.

In practice, without the corresponding frame four-vector one cannot properly define the external charge and current densities in that frame. For the nontrivial transformation of charge and current densities between the comoving and normal frames, see Eq.\ (45) in \cite{Hwang-Noh-2022-Axion-EM}; an example is the Ohm's law where a simple relation between the electric field and current density is imposed in the comoving frame, whereas the EM fields in the normal frame are used in handling Maxwell's equations, see \cite{Baumgarte-Shapiro-2010, Gourgoulhon-2012, Shibata-2015, Hwang-Noh-2022-Axion-EM}.

Equations (\ref{Maxwell-electric-3-4}) and (\ref{Maxwell-magnetic-3-4}) provide the four Maxwell's equations valid in general relativity with exactly the same form as in special relativity. Complications are hidden as our different notations for the EM fields indicate; the other parts of Maxwell's equations are highly complicated. In a short article \cite{Einstein-1916}, Einstein used these coincidences {\it only} to guess the correct form of four-dimensional Maxwell's equations valid in curved spacetime, but he has not defined the EM fields in this manner.

Here is our reason for supporting the normal frame as the proper one to define the EM fields. A timelike unit vector $n_a$ is normal to the hypersurface. It is the four-velocity of an observer instantaneously at rest in the chosen time slice. Therefore, it can be interpreted as an Eulerian observer as its motion follows the hypersurface independently of the coordinates chosen \cite{Smarr-York-1978, Wilson-Mathews-2003, Gourgoulhon-2012}. The coordinate observer in Appendix \ref{sec:coordinate} follows the spatial coordinate. 

To resolve the issue observationally, we may have to further employ the Fermi normal coordinate (FNC) \cite{Manasse-Misner-1963}. The FNC is a specially constructed local inertial frame using geodesics and parallel transport. The local measurement is made by an observer moving along her world line in a given external metric. As we considered the most general metric perturbation without imposing the gauge condition, Maxwell's equations in Eqs.\ (\ref{Maxwell-normal-1})-(\ref{Maxwell-normal-4}) are valid in both the global metric and the FNC. 
The perturbed external metric $h_{ab}$ the observer experience in the FNC are given in terms of the Riemann curvature tensor (tetrad transformed from the one in the global spacetime), which is gauge-invariant to the linear order in Minkowski background, presented in \cite{Manasse-Misner-1963, Fortini-Gualdi-1982, Marzlin-1994, Licht-2004, Rakhmanov-2014, Berlin-2022, Domcke-2022}. Therefore, even in the FNC, defining the appropriate EM fields and charge and current densities by choosing the observer's frame is needed independently.

In the normal frame, the effective ${\bf P}$s and ${\bf M}$s appear in both the homogeneous and inhomogeneous parts of Maxwell's equations, as in Eqs.\ (\ref{Maxwell-normal-1})-(\ref{PM-normal}), and the simple analogy with the axion experiment is broken. However, the structure of Maxwell's equations allows that, by using $\hat {\bf E}$ and $\hat {\bf B}$ in Eq.\ (\ref{EB-electric}) simply as the new parametrization of the ${\bf E}$ and ${\bf B}$, we may benefit the advantage of having $\hat {\bf P}$ and $\hat {\bf M}$ appearing only in the inhomogeneous parts, thus establishing the connection with the axion detection experiments as suggested in \cite{Berlin-2022, Domcke-2022}. That is, we may use Eqs.\ (\ref{Maxwell-electric-1})-(\ref{Maxwell-electric-4}) as the Maxwell's equations to use in the experiment and recover ${\bf E}$ and ${\bf B}$ using Eq.\ (\ref{EB-electric}). Further study in this line is left for future investigation.

%
%
%
\section*{Acknowledgments}

We wish to thank Daniel Braun and Valerie Domcke for fruitful discussion in Moriond Conference. We thank Donghui Jeong and Chan Park for useful discussion. H.N.\ was supported by the National Research Foundation (NRF) of Korea funded by the Korean Government (No.\ 2018R1A2B6002466 and No.\ 2021R1F1A1045515). J.H.\ was supported by IBS under the project code, IBS-R018-D1, and by the NRF of Korea funded by the Korean Government (No.\ NRF-2019R1A2C1003031).

\appendix
%
%
%
\section{Coordinate frame}
                                     \label{sec:coordinate}

In handling Maxwell's equations in curved spacetime, the coordinate frame is often used in the literature \cite{deFelice-1971, Spengler-2023}. The coordinate frame takes $\bar n^i \equiv d x^i/d x^0 \equiv 0$ with the observer at rest in the spatial coordinate \cite{Wilson-Mathews-2003,Baumgarte-Shapiro-2010, Baumgarte-Shapiro-2021}, thus
\bea
   & & \bar n_i = h_{0i}, \quad
       \bar n_0 = - 1 - {1 \over 2} h^0_0,
   \nonumber \\
   & &
       \bar n^i \equiv 0, \quad
       \bar n^0 = 1 - {1 \over 2} h^0_0,
   \label{bar-n_a}
\eea
For the EM fields and current density in the coordinate frame, we set
\bea
   & & \widetilde B_i \equiv \bar B_i, \quad
       \widetilde B_0 = 0,
   \nonumber \\
   & &
       \widetilde B^i = \bar B^i - h^{ij} \bar B_j, \quad
       \widetilde B^0 = h^i_0 \bar B_i,
\eea
and similarly for $\bar E_a$ and $\bar j_a$ with $\widetilde E_i \equiv \bar E_i$ and $\widetilde j_i \equiv \bar j_i$; indices of $\bar B_i$, $\bar E_i$, and $\bar j_i$ are raised and lowered using $\delta_{ij}$ and its inverse. Equations (\ref{Fab-cov}) and (\ref{J^a}) give
\bea
   & & \widetilde F_{0i}
       = - \left( 1 + {1 \over 2} h^0_0 \right) \bar E_i,
       \quad
       \widetilde F_{ij} = h_{0i} \bar E_j - h_{0j} \bar E_i
   \nonumber \\
   & & \qquad
       + \eta_{ijk} \left[ \left( 1
       + {1 \over 2} h^\ell_\ell \right) \bar B^k
       - h^{k\ell} \bar B_\ell \right],
   \nonumber \\
   & & \widetilde J^0
       = \bar \varrho c
       \left( 1 - {1 \over 2} h^0_0 \right)
       + h^i_0 \bar j_i, \quad
       \widetilde J^i = \bar j^i - h^{ij} \bar j_j.
   \label{Fab-coordinate}
\eea
Maxwell's equations follow from Eq.\ (\ref{Maxwell-tensor-eqs}) as
\bea
   & & ( \bar E^i + \bar P_{\rm E}^i )_{,i}
       = \left( 1 + {1 \over 2} h^i_i \right)
       \bar \varrho
       + {1 \over c} h^i_0 \bar j_i,
   \label{Maxwell-coordinate-1} \\
   & &
       ( \bar E^i + \bar P_{\rm E}^i )_{,0}
       - \eta^{ijk} \nabla_j
       ( \bar B_k- \bar M^{\rm E}_k )
   \nonumber \\
   & & \qquad
       =
       - {1 \over c} \left( 1 + {1 \over 2} h^j_j
       + {1 \over 2} h^0_0 \right) \bar j^i
       + {1 \over c} h^{ij} \bar j_j,
   \label{Maxwell-coordinate-2} \\
   & & ( \bar B^i + \bar P_{\rm B}^i )_{,i} = 0,
   \label{Maxwell-coordinate-3} \\
   & & ( \bar B^i + \bar P_{\rm B}^i
       )_{,0}
       + \eta^{ijk} \nabla_j ( E_k - M^{\rm B}_k )
       = 0,
   \label{Maxwell-coordinate-4}
\eea
where the effective ${\bf P}$s and ${\bf M}$s caused by the metric are
\bea
   & & \hskip -1cm \bar P_{\rm E}^i
       \equiv {1 \over 2} h^j_j \bar E^i - \bar h^{ij} E_j
       - \eta^{ijk} h_{0j} \bar B_k, \quad
       \bar M_{\rm E}^i
       \equiv {1 \over 2} h_{00} \bar B^i,
   \nonumber \\
   & & \hskip -1cm \bar P_{\rm B}^i
       \equiv {1 \over 2} h^j_j \bar B^i - h^{ij} \bar B_j
       + \eta^{ijk} h_{0j} \bar E_k, \quad
       \bar M_{\rm B}^i
       \equiv {1 \over 2} h_{00} \bar E^i.
   \label{PM-coordinate}
\eea
We note that $\bar \varrho$ and $\bar j_i$ here differ from $\overline \varrho$ and $\overline j_i$ defined in Eqs.\ (\ref{Maxwell-normal-1}) and (\ref{Maxwell-normal-2}). By comparing Eqs.\ (\ref{Fab-normal}) and (\ref{Fab-coordinate}), we have
\bea
   & & \bar E_i = E_i + \eta_{ijk} h^j_0 B^k, \quad
       \bar B_i = B_i - \eta_{ijk} h^j_0 E^k,
   \nonumber \\
   & & \bar \varrho = \varrho - {1 \over c} h^i_0 j_i,
       \quad
       \bar j_i = j_i - \varrho c h_{0i}.
\eea

We propose the normal frame as the physical one because it corresponds to the frame of an Eulerian observer. As both frames are introduced in covariant manner, the charge and current densities are properly defined and any frame is fine mathematically.

%
%
%
\section{Gauge transformation}
                                     \label{sec:GT}

Under the gauge transformation, $\widehat x^a = x^a + \xi^a$, we have $\widehat h_{ab} = h_{ab} - \xi_{a,b} - \xi_{b,a}$ to the linear order and $R^a_{\;\;bcd}$ in Eq.\ (\ref{metric-pert}) is gauge-invariant. Using the gauge transformation properties of $F_{ab}$ and $J_a$,
\bea
   F_{ab} (x^e) = {\partial \widehat x^c \over \partial x^a}
       {\partial \widehat x^d \over \partial x^b}
       \widehat F_{cd} (\widehat x^e), \quad
       J_a (x^e) = {\partial \widehat x^b \over \partial x^a}
       \widehat J_b (\widehat x^e),
\eea
to the second order in perturbations, we have
\bea
   & & \widehat B^i = B^i - B^i_{\;\;,0} \xi^0
       - \eta^{ijk} E_j \xi^0_{\;\;,k}
       - B^i_{\;\;,j} \xi^j - B_j \xi^{j,i},
   \nonumber \\
   & & \widehat E^i = E^i - E^i_{\;\;,0} \xi^0
       + \eta^{ijk} B_j \xi^0_{\;\;,k}
       - E^i_{\;\;,j} \xi^j - E_j \xi^{j,i},
   \nonumber \\
   & & \widehat \varrho = \varrho
       - \varrho_{, 0} \xi^0
       + {1 \over c} j^i \xi^0_{\;\;,i}
       - \varrho_{, i} \xi^i,
   \nonumber \\
   & & \widehat j^i = j^i - j^i_{\;\;,0} \xi^0
       + \varrho c \xi^{0,i}
       - j^i_{\;\;,j} \xi^j - j_j \xi^{j,i},
\eea
where the index of $\xi^a$ can be raised and lowered using $\eta_{ab}$ and its inverse. Using these gauge transformation properties, we can show that all the Maxwell's equations in this work are gauge-invariant.

%
%


\end{document}